\documentclass[twocolumn]{aastex631}
\accepted{ApJ}

\shorttitle{Molecular jet at earliest phase}
\shortauthors{Dutta et al.}
\graphicspath{{./}{figures/}}

\begin{document}

\title{ALMA Survey of Orion Planck Galactic Cold Clumps (ALMASOP): Evidence for a Molecular Jet Launched at an Unprecedented Early Phase of Protostellar evolution}

\correspondingauthor{Somnath Dutta}
\email{sdutta@asiaa.sinica.edu.tw, cflee@asiaa.sinica.edu.tw}

\author[0000-0002-2338-4583]{Somnath Dutta}
\affiliation{Institute of Astronomy and Astrophysics, Academia Sinica, Roosevelt Rd, Taipei 10617, Taiwan, R.O.C.}

\author[0000-0002-3024-5864]{Chin-Fei Lee}
\affiliation{Institute of Astronomy and Astrophysics, Academia Sinica, Roosevelt Rd, Taipei 10617, Taiwan, R.O.C.}

\author[0000-0001-9304-7884]{Naomi Hirano}
\affiliation{Institute of Astronomy and Astrophysics, Academia Sinica, Roosevelt Rd, Taipei 10617, Taiwan, R.O.C.}

\author[0000-0002-5286-2564]{Tie Liu}
\affiliation{Shanghai Astronomical Observatory, Chinese Academy of Sciences, 80 Nandan Road, Shanghai 200030, China}
\affiliation{Key Laboratory for Research in Galaxies and Cosmology, Chinese Academy of Sciences, 80 Nandan Road, Shanghai 200030 People’s Republic of China}

\author[0000-0002-6773-459X]{Doug Johnstone}
\affiliation{NRC Herzberg Astronomy and Astrophysics, 5071 West Saanich Rd., Victoria, BC V9E 2E7, Canada}
\affiliation{Department of Physics and Astronomy, University of Victoria, Victoria, BC V8P 1A1, Canada}

\author[0000-0003-4603-7119]{Sheng-Yuan Liu}
\affiliation{Institute of Astronomy and Astrophysics, Academia Sinica, Roosevelt Rd, Taipei 10617, Taiwan, R.O.C.}

\author[0000-0002-8149-8546]{Ken'ichi Tatematsu}
\affil{Nobeyama Radio Observatory, National Astronomical Observatory of Japan, 
National Institutes of Natural Sciences, 
462-2 Nobeyama, Minamimaki, Minamisaku, Nagano 384-1305, Japan}
\affiliation{Department of Astronomical Science,
SOKENDAI (The Graduate University for Advanced Studies),
2-21-1 Osawa, Mitaka, Tokyo 181-8588, Japan}

\author{Paul F. Goldsmith}
\affiliation{Jet Propulsion Laboratory, California Institute of Technology, 4800 Oak Grove Drive, Pasadena, CA 91109, USA}

\author[0000-0002-4393-3463]{Dipen Sahu}
\affiliation{Institute of Astronomy and Astrophysics, Academia Sinica, Roosevelt Rd, Taipei 10617, Taiwan, R.O.C.}

\author{Neal J. Evans}
\affiliation{Department of Astronomy The University of Texas at Austin 2515 Speedway, Stop C1400 Austin, TX 78712-1205, USA}

\author[0000-0002-7125-7685]{Patricio Sanhueza} 
\affiliation{National Astronomical Observatory of Japan, National Institutes of Natural Sciences, 2-21-1 Osawa, Mitaka, Tokyo 181-8588, Japan}
\affiliation{Department of Astronomical Science,
SOKENDAI (The Graduate University for Advanced Studies),
2-21-1 Osawa, Mitaka, Tokyo 181-8588, Japan}

\author{Woojin Kwon}
 \affil{Department of Earth Science Education, Seoul National University, 1 Gwanak-ro, Gwanak-gu, Seoul 08826, Republic of Korea}
 \affil{SNU Astronomy Research Center, Seoul National University, 1 Gwanak-ro, Gwanak-gu, Seoul 08826, Republic of Korea}

\author{Sheng-Li Qin}
\affiliation{Department of Astronomy, Yunnan University, and Key Laboratory of Particle Astrophysics of Yunnan Province, Kunming, 650091, People's Republic of China}

\author{Manash Ranjan Samal}
\affiliation{Physical Research Laboratory, Navrangpura, Ahmedabad, Gujarat 380009, India}

\author[0000-0003-2384-6589]{Qizhou Zhang}
\affiliation{Center for Astrophysics | Harvard \&amp; Smithsonian, 60 Garden Street, Cambridge, MA 02138, USA}

\author[0000-0003-2412-7092]{Kee-Tae Kim}
\affil{Korea Astronomy and Space Science Institute (KASI), 776 Daedeokdae-ro, Yuseong-gu, Daejeon 34055, Republic of Korea}
\affil{University of Science and Technology, Korea (UST), 217 Gajeong-ro, Yuseong-gu, Daejeon 34113, Republic of Korea}

\author[0000-0001-8385-9838]{Hsien Shang}
\affiliation{Institute of Astronomy and Astrophysics, Academia Sinica, Roosevelt Rd, Taipei 10617, Taiwan, R.O.C.}

\author{Chang Won Lee}
\affil{Korea Astronomy and Space Science Institute (KASI), 776 Daedeokdae-ro, Yuseong-gu, Daejeon 34055, Republic of Korea}
\affil{University of Science and Technology, Korea (UST), 217 Gajeong-ro, Yuseong-gu, Daejeon 34113, Republic of Korea}

\author{Anthony Moraghan}
\affiliation{Institute of Astronomy and Astrophysics, Academia Sinica, Roosevelt Rd, Taipei 10617, Taiwan, R.O.C.}

\author{Kai-Syun Jhan}
\affiliation{Institute of Astronomy and Astrophysics, Academia Sinica, Roosevelt Rd, Taipei 10617, Taiwan, R.O.C.}

\author[0000-0003-1275-5251]{Shanghuo Li}
\affiliation{Korea Astronomy and Space Science Institute (KASI), 776 Daedeokdae-ro, Yuseong-gu, Daejeon 34055, Republic of Korea}

\author{Jeong-Eun Lee} 
\affil{School of Space Research, Kyung Hee University, Yongin-Si, Gyeonggi-Do 17104, Republic of Korea}

\author{Alessio Traficante}
\affil{IAPS-INAF, via Fosso del Cavaliere 100, I-00133, Rome, Italy}

\author[0000-0002-5809-4834]{Mika Juvela}
\affiliation{Department of Physics, P.O.Box 64, FI-00014, University of Helsinki, Finland}

\author[0000-0002-9574-8454]{Leonardo Bronfman}
\affil{Departamento de Astronomía, Universidad de Chile, Casilla 36-D, Santiago, Chile}

\author{David Eden}
\affiliation{Astrophysics Research Institute, Liverpool John Moores University, IC2, Liverpool Science Park, 146 Brownlow Hill, Liverpool, L3 5RF, UK}

\author[0000-0002-6386-2906]{Archana Soam}
\affil{SOFIA Science Center, Universities Space Research Association, NASA Ames Research Center, Moffett Field, California 94035, USA}

\author[0000-0002-3938-4393]{Jinhua He}
\affil{Yunnan Observatories, Chinese Academy of Sciences, 396 Yangfangwang, Guandu District, Kunming, 650216, P. R. China}
\affiliation{Chinese Academy of Sciences South America Center for Astronomy, National Astronomical Observatories, CAS, Beijing 100101, China}
\affiliation{Departamento de Astronom{\'i}a, Universidad de Chile, Casilla 36-D, Santiago, Chile}

\author[0000-0003-3343-9645]{Hong-li Liu}
\affil{Department of Astronomy, Yunnan University, Kunming 650091, People’s Republic of China}

\author[0000-0002-4336-0730]{Yi-Jehng Kuan}
\affiliation{Department of Earth Sciences, National Taiwan Normal University, Taipei, Taiwan, R.O.C. \& Institute of Astronomy and Astrophysics, Academia Sinica, Roosevelt Rd, Taipei 10617, Taiwan, R.O.C.}

\author[0000-0002-8898-1047]{Veli-Matti Pelkonen}
\affiliation{Institut de Ci\`{e}ncies del Cosmos, Universitat de Barcelona, IEEC-UB, Mart\'{i} i Franqu\`{e}s 1, E08028 Barcelona, Spain}

\author{Qiuyi Luo}
\affiliation{Shanghai Astronomical Observatory, Chinese Academy of Sciences, 80 Nandan Road, Shanghai 200030, China}

\author[0000-0003-0537-5461]{Hee-Weon Yi}
\affiliation{Korea Astronomy and Space Science Institute (KASI), 776 Daedeokdae-ro, Yuseong-gu, Daejeon 34055, Republic of Korea}

\author{Shih-Ying Hsu}
\affiliation{Institute of Astronomy and Astrophysics, Academia Sinica, Roosevelt Rd, Taipei 10617, Taiwan, R.O.C.}



\begin{abstract}
Protostellar outflows and jets play a vital role in star formation as they carry away excess angular momentum from the inner disk surface, allowing the material to be transferred toward the central protostar. Theoretically,  low velocity and poorly collimated outflows appear from the beginning of the collapse, at the first hydrostatic core (FHSC) stage. With growing protostellar core mass, high-density jets are launched which entrain an outflow from the infalling envelope.  Until now, molecular jets have been  observed at high velocity ($\gtrsim$ 100 km/s) in early Class\,0 protostars.  We, for the first time, detect a dense molecular jet in SiO emission with small-velocity ($\sim$ 4.2 km\,s$^{-1}$, deprojected $\sim$ 24 km\,s$^{-1}$) from source G208.89-20.04Walma (hereafter, G208Walma) using ALMA Band\,6 observations. This object has some characteristics of FHSCs, such as a small outflow/jet velocity, extended 1.3\,mm continuum emission, and N$_2$D$^+$ line emission. Additional characteristics, however, are typical of early protostars: collimated outflow and  SiO jet. The full extent of the outflow corresponds to a dynamical time scale of $\sim$ 930$^{+200}_{-100}$ years. The spectral energy distribution also suggests a very young source having an upper limit of T$_{bol}$ $\sim$ 31 K and L$_{bol}$ $\sim$ 0.8 L$_\sun$. We conclude that G208Walma is likely in the transition phase from FHSC to protostar, and the molecular jet has been launched within a few hundred years of initial collapse. Therefore, G208Walma may be the earliest object discovered in the protostellar phase with a molecular jet. 
\end{abstract}



\section{Introduction} \label{sec:introduction}

In the standard theory of low-mass star formation \citep[][]{1987ARA&A..25...23S}, a prestellar core contracts quasistatically into a dense core \citep[e.g.,][]{1969MNRAS.145..271L,1979PASJ...31..697N,1987ASIC..210..173L},
and dynamical inside-out collapse follows the formation of a singular isothermal sphere \citep[][]{1977ApJ...214..488S}. In the presence of a magnetic field, an extended pseudo-disk (appearing as a flattened envelope) of a few thousand AU forms perpendicular to the field threading the core \citep[][]{1993ApJ...417..220G} during this transition phase from prestellar core to protostar formation \citep[][]{1996ApJ...472..211L,Allen03}. Very slow and poorly collimated outflows, with velocity $<$ 10 km/s, can emerge from cores having initial rotation \citep[][]{Allen03,2008ApJ...681.1356M,2008A&A...477....9H}.
In some Magneto hydrodynamic (MHD)  simulations, a transient object called the first (opaque) core (later often dubbed the first hydrostatic core; FHSC) can appear at the center of the pseudo-disk \citep[e.g.,][]{2002ApJ...575..306T,2010MNRAS.409L..39C}.
The candidate FHSC may have an observed spectral energy distribution (SED) intermediate between those of prestellar cores and Class\,0 sources \citep[][]{2020MNRAS.499.4394M}.

Protostellar outflows and jets are among the most intriguing phenomena associated with the accretion process during star formation since they are believed to carry away excess angular momentum from the disk surface, therefore, allowing material to fall onto the central protostar  \citep[][]{2016ARA&A..54..491B,2020A&ARv..28....1L}. Theoretically, an  outflow of low-density, extended material is expected to appear at the earliest phase of the collapse process, i.e., the first core formation stage \citep[][]{1969MNRAS.145..271L,2014ApJ...796L..17M}. In contrast, high-density jets are expected to be launched a few hundred years after the initial collapse, possibly from the second core (protostar). 
With time, the fast jet will catch up with the tips of the previous, slow outflow \citep[][]{2002ApJ...575..306T,2014ApJ...796L..17M,2019ApJ...876..149M}. 
To date, molecular jets  with very high jet velocities, $>$ 100 km\,s$^{-1}$ and  mass-loss rates, $\gtrsim$ 10$^{-6}$ M$_\sun$\,yr$^{-1}$, have been observed around Class\,0 protostars  \citep[][]{2020A&ARv..28....1L}.
While FHSC candidates and their associated outflows have been reported \citep[e.g.,][]{2012ApJ...751...89C,2020MNRAS.499.4394M}, the start of the fast jet-launching phase during  protostellar collapse, i.e., the intermediate stage between the FHSC and protostar formation, remains less explored observationally.

A sample of extremely cold dense cores has been observed as a part of the ALMA Survey of Orion Planck Galactic Cold Clumps \citep[ALMASOP;][]{2020ApJS..251...20D}, which opens the opportunity to detect the earliest stage of protostars and investigate the initial condition of protostellar evolution. In this manuscript, we present the detection of a dense molecular SiO jet at an unprecedentedly early protostellar stage, possibly the FHSC-to-protostellar transition phase, in the dense core G208.89-20.04Walma (hereafter, G208Walma), located at a distance of $\sim$ 400 pc \citep[an average distance of Orion star-forming regions,][]{2018AJ....156...84K}. This is potentially the first detection of an SiO jet at such an early  evolutionary stage.  
In section 2, we describe the observations used. Section 3 deals with the outflow, jet, and envelope properties. The evolutionary phase of G208Walma is discussed in section 4, with the conclusion in section 5.

\begin{figure*}
\fig{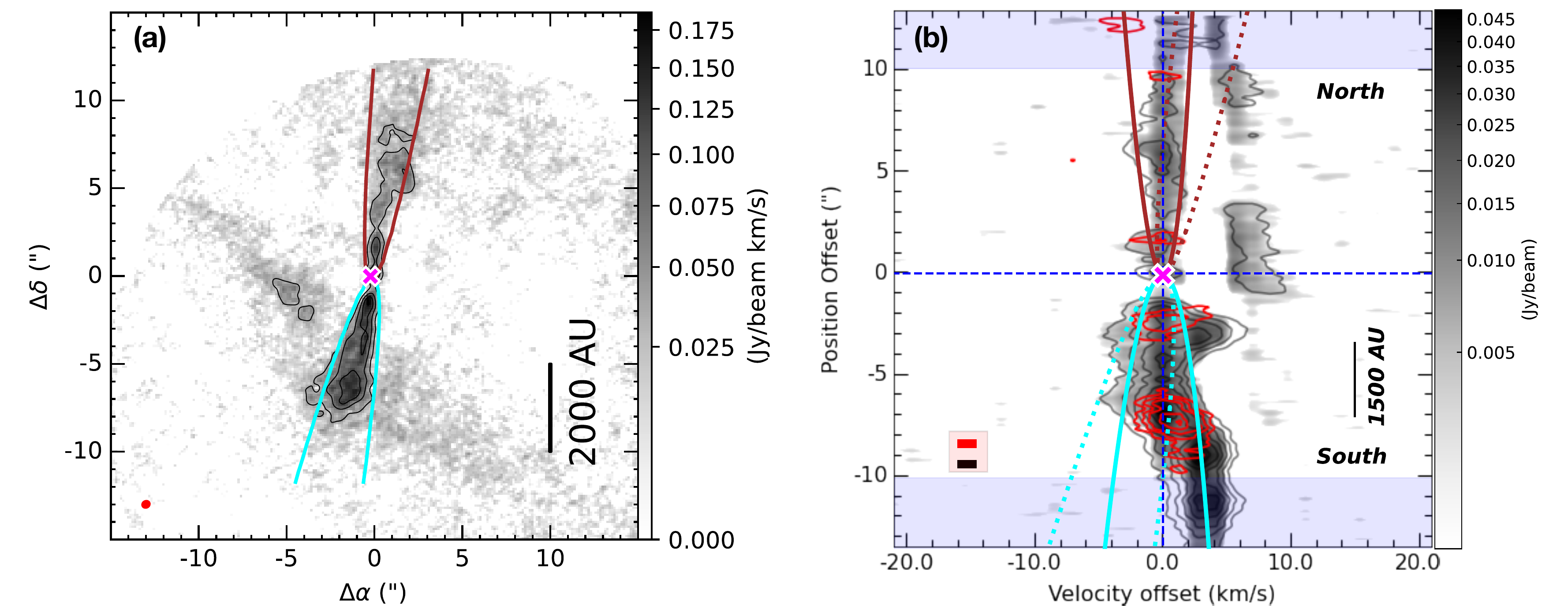}{0.98\textwidth}{}
\caption{(a) High-resolution ALMA  map of G208Walma in CO  emission integrated over 2.8 -- 8.4 km\,s$^{-1}$. The contours are over plotted at 3$\times$(1, 2, 3, 6, 9)$\sigma$, where sensitivity $\sigma$ = 12 mJy\,beam$^{-1}$\,km\,s$^{-1}$.  The synthesized beam size $\sim$ 0$\farcs$41 $\times$ 0$\farcs$34 is shown in the lower left. CO emission is extended in the North-East direction with a position angle of $\sim$ 80$\degr$. The cyan and red parabolas are best fits for  proportionality constant C = 3.3 $\rm arcsec^{-1}$ and 5.0 $\rm arcsec^{-1}$, respectively (see section \ref{sec:CO_SiO_PV} for details).
(b) Position Velocity (PV) diagram along the jet axis of $^{12}$CO(2$-$1) emission. The black contours are at 3$\times$(1, 2, 3, 4, 5, 6)$\sigma$, where $\sigma$ = 1.0 mJy\,beam$^{-1}$.  The SiO contours are over plotted in red at (2.5, 3.5, 4.5, 6.5, 7.5, 18)$\sigma$, where $\sigma$ = 0.9 mJy\,beam$^{-1}$. The beam sizes are shown in lower left for  CO (in black) and SiO (in red) emission. The solid and dotted parabolas are for $i_a$ $\sim$ 1$\degr$ and 10$\degr$, respectively. Cyan and red represent the Southern and Northern lobe, respectively. 
The blue dashed straight lines, representing the zero-axis, cross at the source position (magenta cross).  The emission within the blue-shaded region at $>$ 10$\arcsec$ is possibly tracing ambient material (see text for details). Linear scale bars are shown in both the panels.
} 
\label{fig:CO_SiO_PV}
\end{figure*}

\begin{figure*}
\fig{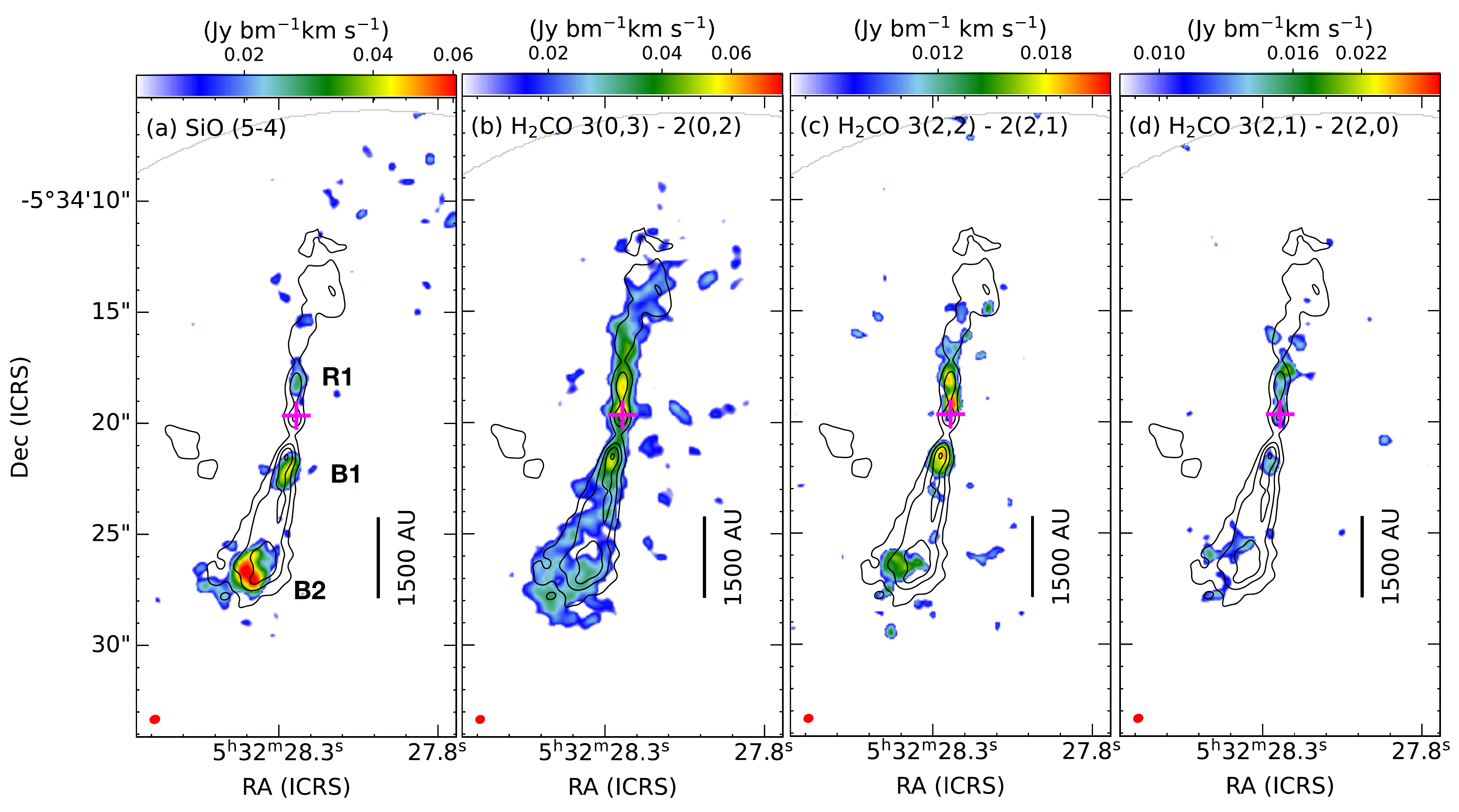}{0.95\textwidth}{}
\caption{
 High-resolution ALMA integrated maps (moment 0) of G208Walma in (a) SiO\,J=5$-$4  integrated over velocity 2.8 -- 8.4 km\,s$^{-1}$ with sensitivity of $\sim$ 10 mJy\,beam$^{-1}$\,km\,s$^{-1}$,  (b) H$_2$CO\,3( 0, 3)$-$2( 0, 2) with sensitivity of $\sim$ 9.4 mJy\,beam$^{-1}$\,km\,s$^{-1}$ (c) H$_2$CO\,3( 2, 2)$-$2( 2, 1) with sensitivity of $\sim$ 6.5 mJy\,beam$^{-1}$\,km\,s$^{-1}$, and   (d) H$_2$CO\,3( 2, 1)$-$2( 2, 0) with sensitivity of $\sim$ 6 mJy\,beam$^{-1}$\,km\,s$^{-1}$ at spatial resolution of 140 AU. All H$_2$CO transitions are integrated over velocity 7.0 -- 8.4 km\,s$^{-1}$.
$^{12}$CO(2-1) contours have the same meaning as  Figure \ref{fig:CO_SiO_PV}a. The prominent Knot-like structures are marked as B1, B2 in the Southern lobe and R1 in the Northern lobe.
The continuum peak position is marked with a magenta cross in all panels. The beam sizes are shown in red on the lower left (typically $\sim$ 0$\farcs$41 $\times$ 0$\farcs$34). The linear scale bars are shown in all  the panels.
}
\label{fig:SiO_CO_H2CO_jet}
\end{figure*}

\section{Observations}\label{sec:observations}
G208Walma was observed with ALMA as a part of the ALMASOP survey of extremely young dense cores (Project ID:2018.1.00302.S; PI: Tie Liu) in Band\,6 
\citep[see][for more details on ALMASOP]{2020ApJS..251...20D}. This paper discusses 1.3\,mm dust continuum, C$^{18}$O\,(2$-$1), N$_2$D$^+$\,(3$-$2), CO\,(2$-$1), SiO\,(5$-$4) and three  H$_2$CO transitions: 3(0,3)$-$2(0,2), 3(2,2)$-$2(2,1), 3(2,1)$-$2(2,0). The calibration of the acquired visibilities was performed with the standard pipeline in CASA 5.4 \citep{2007ASPC..376..127M}. Two sets of continuum and line-cubes were generated with the TCLEAN task: (i) combining visibilities of three configurations (i.e., TM1+TM2+ACA), which produces an image of a typical synthesized beam size $\sim$ 0$\farcs$41 $\times$ 0$\farcs$34 ($-$69$\degr$), and (ii) only using the 7-m ACA configuration to produce a typical synthesized beam size $\sim$ 7$\farcs$92 $\times$ 4$\farcs$42 ($-$80$\degr$) for continuum  and $\sim$ 8$\farcs$42 $\times$ 4$\farcs$67 ($-$79$\degr$) for line cubes. From here on, we refer to the first case as high-resolution and the latter case as low-resolution.  We applied a robust weighting factor of R$_w$ = $+$2.0 (natural weighting) for the 1.3-mm ACA continuum, which is adequate to achieve maximum continuum flux for dust mass estimation. For the remaining cases, we utilized R$_w$ = $+$0.5. The continuum maps were generated with a threshold of 3\,$\sigma$, where $\sigma$ is the  theoretical sensitivity. 
The line-cubes have a velocity resolution of $\sim$ 1.4 km\,s$^{-1}$, and typical sensitivity of $\sim$ 3 mJy\,beam$^{-1}$ at high-resolution and 30 mJy\,beam$^{-1}$ at low-resolution. More details on the data analyses are presented by \citet[][]{2020ApJS..251...20D}. 

\begin{deluxetable}{l@{\extracolsep{2pt}}c@{\extracolsep{10pt}}c@{\hskip 1pt}c@{\extracolsep{1pt}}
}[h]
\tablecaption{Physical properties of the Jet and Outflow}\label{tab:jet_outflow}
\tablewidth{0pt}
\tablehead{
\colhead{Parameter (Unit)} & \colhead{South} & \colhead{North} & \colhead{Total} 
}
\startdata
\multicolumn{4}{c}{Jet}\\
\cline{1-4}
V$_{\rm obs,out}$$^*$ (km\,s$^{-1}$) & 4.2  & 4.2 & $-$\\
N$_{\rm CO}$ (10$^{16}$ cm$^{-2}$) & 2.30 & 2.02 & $-$\\
\.{M}$_{\rm j}$ (10$^{-7}$ M$_\sun$\,yr$^{-1}$) & 1.01 & 0.95 & 1.96\\
L$_{\rm mech}$ (10$^{-3}$ L$_\sun$) & 1.03 & 0.75 & 1.78\\
\cline{1-4}
N$_{\rm SiO}$ (10$^{13}$ cm$^{-2}$) & 2.7 & 1.8 &$-$\\
X[SiO/CO] (10$^{-4}$)     &  5.8   & 4.2 &$-$\\
\cline{1-4}
\multicolumn{4}{c}{Outflow}\\
Size$^\dagger$ (10$^3$ au)    & 4.5 & 5.0 & $-$\\
t$_{dyn}$\,$^\dagger$ (yr) & 890    &  970 & $-$\\
F$_{\rm  CO}$ (10$^{-8}$ M$_\sun$km s$^{-1}$/yr)  &  0.5   & 0.3 & 0.8\\
\enddata
\tablecomments{$^*$ The outflow/jet velocity (V$_{\rm obs,out}$) is estimated to be 4.2 km\,s$^{-1}$. The deprojected jet velocity, V$_j$ = 24 km\,s$^{-1}$, was obtained for an inclination angle of 10$\degr$.\\
$^\dagger$ Estimated based on full outflow extension as observed at 7-m ACA primary beam.
}
\end{deluxetable}

\section{Results}\label{sec:results}

\subsection{Physical Structure of CO  outflow}\label{sec:CO_SiO_PV}
The $^{12}$CO emission map integrated over velocity range 2.8 -- 8.4 km\,s$^{-1}$ and the position-velocity (PV) diagram along the outflow/jet axis are shown in Figure \ref{fig:CO_SiO_PV}a and b, respectively. SiO contours are over-plotted in red in Figure \ref{fig:CO_SiO_PV}b. CO and SiO emissions are extended in the North-East direction with a position angle of $\sim$ 80$\degr$. 
The velocity axis in Figure \ref{fig:CO_SiO_PV}b is the velocity offset from the systemic velocity,  V$_{sys}$ = 7 km\,s$^{-1}$, estimated from the N$_2$D$^+$ emission. Here velocity offset, V$_{\rm off}$ = $|V_{\rm obs}$ -- $V_{\rm sys}|$, where $V_{\rm obs}$ is the observed channel velocity. Visual inspection of the channel maps (Figure \ref{fig:Appendix_CO_channelMap}) and CO/SiO spectra (Figure \ref{fig:Appendix_CO_SiO_spectra}) suggest that emission beyond 4.2 km\,s$^{-1}$ velocity offset is tracing ambient material. In Figure \ref{fig:CO_SiO_PV}b, the emission in the Northern lobe (positive position offset) has two clear isolated parallel components: one around the source velocity and another beyond 4.2 km\,s$^{-1}$ velocity offset, which is likely the ambient material.   Similarly, the Southern lobe (negative position offset) is combined with ambient cloud at $>$ 4.2 km\,s$^{-1}$ velocity offset.  In Figure \ref{fig:CO_SiO_PV}b, the shaded region in blue at the end of each lobe, beyond 10$\arcsec$, is dominated by ambient material in the combined configuration TM1+TM2+ACA maps. Based on SiO and CO emission, the maximum observed velocity offset (V$_{\rm obs,out}$) of the outflow/jet is 4.2 kms$^{-1}$. The PV diagrams and the spectra of CO and SiO do not exhibit any high-velocity component, therefore it is difficult to disentangle the velocity distribution of the outflow, swept up material, and jet components.

Following the simple analytical model by \citet[][]{2000ApJ...542..925L}, the physical structure of the outflow shell in CO emission is described by the equation $z = CR^2$, 
where $R$ is the (cylindrical) radial distance from the outflow axis $z$ and $C$ is a proportionality constant. In  Figure \ref{fig:CO_SiO_PV}a, fitting  a  parabola to the outermost contour of the outflow yields C = 3.3 $\rm arcsec^{-1}$ and 5.0 $\rm arcsec^{-1}$ for the Southern and Northern lobe, respectively. With these $C$ values, we aim to fit parabolas to the respective outflow lobes in the PV diagram (Figure \ref{fig:CO_SiO_PV}b). We are unable to obtain a fit to the PV structures. We display two example fits assuming inclination angles $i_a$ = 1$\degr$ and 10$\degr$ for both the lobes. From the ratio of major and minor axis in 1.3\,mm continuum and N$_2$D$^+$ emission (see section \ref{sec:envelope}), we estimated an $i_a$ $\sim$ 20$\degr$. However geometrically thick emission from young stars might not demonstrate the actual orientation of the disk. Thus, we conclude that the $i_a$ could be in the range 1$-$20$\degr$. We therefore assume an intermediate $i_a$ $\sim$ 10$\degr$ for G208Walma, which is utilized for deprojection of other observed parameters.

\subsection{Detection of a dense molecular jet}\label{sec:outflow_jet}
\begin{figure}
\fig{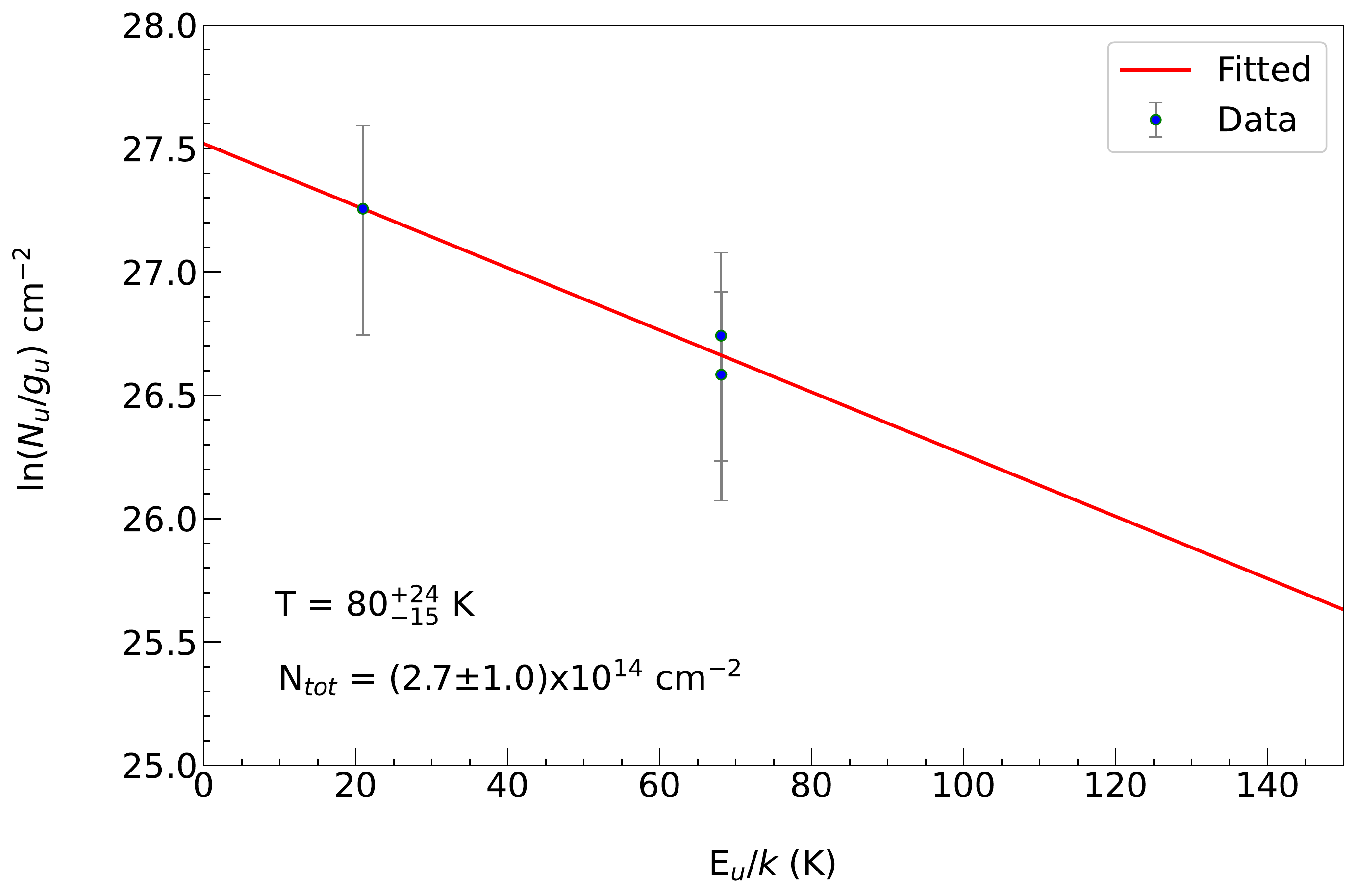}{0.45\textwidth}{}
\caption{
 Rotation diagram for molecular transitions of H$_2$CO. The diagram is derived from the line intensities in the outflow/jet of Figure \ref{fig:SiO_CO_H2CO_jet}b-d, as listed in Table \ref{tab:H2CO_spatalogue}. The error bars indicate the uncertainty in our measurements, which are assumed to be 40\% of the estimated data values. The solid line is a linear fit to the data. The rotational temperature and column density are estimated as $\sim$  80$^{+24}_{-15}$ K and (2.7 $\pm$ 1.00) $\times$ 10$^{14}$ cm$^{-2}$ from the fitting.
}
\label{fig:H2CO_rotational_diagram}
\end{figure}

Figures \ref{fig:SiO_CO_H2CO_jet}a-d show the  high-resolution integrated intensity maps in SiO (5--4) and three H$_2$CO transitions. A rotation diagram using the energy levels of H$_2$CO transitions is shown in Figure \ref{fig:H2CO_rotational_diagram}.
High-resolution integrated intensity CO contours from Figure \ref{fig:CO_SiO_PV}a are over plotted in Figures \ref{fig:SiO_CO_H2CO_jet}a-d for comparison and the low-resolution CO contours are displayed in Figure \ref{fig:cont_N2D_C18O}c.

In section \ref{sec:CO_SiO_PV}, we found that 
the maximum flow velocity in  CO emission is V$_{\rm max,CO}$ $\sim$  4.2 km\,s$^{-1}$ (Table \ref{tab:jet_outflow}) and similarly in SiO, V$_{\rm max,SiO}$ $\sim$  4.2 km\,s$^{-1}$. 
 Both SiO and CO emission exhibit low-velocity components with a very similar velocity range, complicating the separation of the jet component from swept-up material (outflow) based on flow velocity only. Therefore, we define the jet component by considering the density perspective using SiO and CO. 
CO emission above the 3\,$\sigma$ contour is considered as outflow shell, as shown in Figure \ref{fig:CO_SiO_PV}a.
 On the other hand, observations of other sources in SiO (5--4) have shown that it traces the dense shocked gas, and SiO (5--4) emission along the outflow axis mainly implies the high-density jet  \citep[e.g.,][references therein]{2016ARA&A..54..491B,2020A&ARv..28....1L}.  We compute the critical density for SiO (5--4) emission as $\sim$ 2.5$-$3.0 $\times$ 10$^6$ cm$^{-3}$ for jet temperatures between 50 - 300 K, where the Einstein A coefficient is adopted from CDMS database \citep[][]{2001A&A...370L..49M} and collisional rate coefficients from \citep[][]{2018MNRAS.479.2692B}. Such critical densities should be reached only in the jet or knots.  Therefore, despite the low velocity of the outflow/jet material, we consider the SiO (5--4) emission, found along the outflow/jet axis of G208Walma, to indicate the jet component.


A few knot-like structures (B1, B2, R1 in Figure \ref{fig:SiO_CO_H2CO_jet}) are prominent in SiO emission, and are also traced by CO and higher transitions of H$_2$CO emission (Figures \ref{fig:SiO_CO_H2CO_jet}c-d). Two knots closest to the source (B1 and R1 in Figure \ref{fig:SiO_CO_H2CO_jet}a) on the Southern and Northern lobes are possibly part of the jet, whereas the SiO emission at outermost part of Southern lobe (B2 in Figure \ref{fig:SiO_CO_H2CO_jet}a) might be part of the collision zone between the jet/outflow and ambient material (see also Figure \ref{fig:CO_SiO_PV}b,   \ref{fig:cont_N2D_C18O}d, and \ref{fig:Appendix_CO_channelMap}).  The maximum outflow and jet observed velocity, V$_{\rm obs,out}$ = 4.2 km\,s$^{-1}$, based on CO and SiO emission, corresponds to a de-projected jet velocity of V$_{\rm j}\sim$ 24 km\,s$^{-1}$ for an assumed inclination angle, $i_{\rm a}$ =10$\degr$ (see section \ref{sec:CO_SiO_PV} for details). These outflow/jet properties are listed  in Table  \ref{tab:jet_outflow}.

 The jet mass-loss rate \.{M$_j$} was derived using the average CO emission from the two knot-areas close to the source (B1 and R1 in Figure \ref{fig:SiO_CO_H2CO_jet}). Assuming optically thin emission in the jet, we estimate a beam-average CO column density (N$_{\rm CO}$) of $\sim$ 2.16 $\times$ 10$^{16}$ cm$^{-2}$. We assume a high CO excitation temperature of T$_{\rm ex,jet}$ $\sim$ 150\,K, since the jet emission is associated with  internal shocks at high temperature \citep[][]{2004ApJ...603..198G,2007ApJ...659..499L,2010ApJ...713..731L}. N$_{\rm CO}$ is then converted into beam-averaged H$_2$ column density, N$_{{\rm H}_2}$, assuming a CO abundance ratio, X$_{\rm CO}$  = N$_{\rm CO}$/N$_{{\rm H}_2}$ = 4 $\times$ 10$^{-4}$ \citep[][]{1991ApJ...373..254G}. However, we note that this ratio could be as small as $\sim$ 10$^{-4}$ \citep[][]{2010ApJ...717...58H,2015A&A...576A.109Y}, hence the measured N$_{{\rm H}_2}$ should be considered a lower limit. Using equation \ref{equ:mass_equation} from Appendix \ref{sec:appendix_results_jetmassloss_rate}, the total jet mass-loss rate (\.{M}$_{\rm j}$) is estimated to be $\sim$ 1.96 $\times$ 10$^{-7}$ M$_\sun$\,yr$^{-1}$, considering the deprojected velocity. 
A very small mechanical luminosity (L$_{\rm mech}$ $\sim$ 1.78 $\times$ 10$^{-3}$  L$_\sun$) is found. Similar values of \.M$_{\rm j}$ and L$_{\rm mech}$ were also observed toward other Class\,0 and Class\,I protostars with SiO jets \citep[e.g.,][]{2021A&A...648A..45P,2021arXiv211014035D}.

For the low-density outflow, we assume a smaller mean specific excitation temperature of T$_{\rm ex,out}$ = 50\,K with a smaller abundance ratio, X$_{\rm CO}$ $\sim$ 10$^{-4}$, than that for the jet  \citep[][]{2015A&A...576A.109Y,2022ApJ...925...11D}. We use a lower temperature here since the low-density outflow is coasting into  ambient material of cooler environment. Using equation \ref{equ:force_equation} from Appendix \ref{sec:Appendix_result_forcemomentum}, we estimate a total outflow force (F$_{\rm CO}$) of $\sim$ 0.8 $\times$ 10$^{-8}$  M$_\sun$km\,s$^{-1}$/yr 
for the entire CO emission.  We note that F$_{\rm CO}$ for each channel is proportional to the velocity of that channel, V$_{\rm k}$. G208Walma has velocity dispersion on both sides of the systemic velocity (V$_{\rm sys}$ = 7.0 km\,$^{-1}$), which makes it difficult to estimate the exact V$_{\rm k}$ from our low-velocity resolution ($\sim$ 1.4 km\,s$^{-1}$) observations. High velocity-resolution observations, therefore, might alter the outflow force measurement by a factor of 2--4. Furthermore, for G208Walma, the outflow emission is collimated and mostly covers the emitting area of the jet, where T$_{\rm ex,out}$ is likely higher. Therefore, 
the estimated F$_{\rm CO}$ should be a lower limit. 

Low-resolution ACA maps offer maximum field-of-view, covering the full extent of the outflow lobes (Figure \ref{fig:cont_N2D_C18O}c). From this map, we estimate an average length for the lobes of $\sim$ 12$\arcsec$ ($\sim$ 4800 au), which corresponds to an average deprojected dynamical time scale of t$_{\rm dyn}$ $\sim$ 930$^{+200}_{-100}$ yr (Table  \ref{tab:jet_outflow}) for $i_{\rm a}$ = 10$\degr$. Here, the error bars are derived based on the uncertainty in the before-projection lobe length estimation. A change of $i_{\rm a}$ $\sim$ 5$\degr$ can alter this value by $\sim$ 50\%.

Similar to the CO analysis, we estimate a mean SiO column density (N$_{\rm SiO}$) of 4.5 $\times$ 10$^{13}$ cm$^{-2}$ from the mean integrated SiO emission with the assumption of optically thin emission and a T$_{\rm ex,jet}$ of 150 K within the jet.  The average SiO/CO abundance ratio, X[SiO/CO] is $\sim$ 5.0 $\times$ 10$^{-4}$. Such high SiO abundance has been previously observed during the a very early stage of the protostellar collapse \citep[][]{2010ApJ...717...58H,2020A&A...636A..60T,2021A&A...648A..45P,2021arXiv211014035D}.  .

\subsection{H$_2$CO along the jet axis}\label{sec:H2CO_outflow_jet}
H$_2$CO commonly traces both outflow and disk/envelope emission.  If carbon molecules dissociate due to high temperatures ($>$ 20\,K), H$_2$CO molecules synthesize around protostars \citep[][]{2017ApJ...839...43O}.  
Figure \ref{fig:SiO_CO_H2CO_jet}b-d displays three transitions of H$_2$CO. From the maps, it is not clear whether H$_2$CO traces either the disk or envelope for this source, although it efficiently traces the outflow lobes. The lower transitions are excited at low temperatures, which can trace the entire outflow lobe (e.g., Figure \ref{fig:SiO_CO_H2CO_jet}b). The higher transitions are more concentrated toward the dense knot structures along with the jet (e.g., Figure \ref{fig:SiO_CO_H2CO_jet}d). 
%

The average observed integrated flux at the knot-positions for the three transitions of H$_2$CO are listed in Table \ref{tab:H2CO_spatalogue}. The Einstein coefficients (A$_{ul}$) and upper energy levels (E$_u$) are obtained from Spatalogue (https://splatalogue.online//). We derive the mean excitation temperature and column density in the outflow/jet using the $^{12}$CO rotational temperature diagram. Here we assume optically thin emission and the same excitation temperature for all transitions. Figure \ref{fig:H2CO_rotational_diagram} displays the column density per statistical weight (N$^{\rm thin}_{\rm u}$/$\rm {g_u}$) as a function of the upper energy level (E$_{\rm u}$) of the lines (see Table \ref{tab:H2CO_spatalogue} for parameters).  Here N$^{\rm thin}_{\rm u}$ = (8$\pi$\,$k\nu^2$/$hc^3A_{ul}$)I, where the integrated line intensity is I = $\int$T$_B$d$v$ for  brightness temperature T$_B$. The best fit rotational temperature is derived as T$_{\rm rot}$ = 80$^{+24}_{-15}$\,K.  Assuming this temperature represents the excitation temperature of the jet/outflow,  we estimated a total column density, N$^{\rm tot}_{\rm {H_2CO}}$  $\sim$ (2.7 $\pm$ 1.0) $\times$ 10$^{14}$ cm$^{-2}$.  
Note that the H$_2$CO transitions may trace different components in different transitions, such as the outflow in the lower transitions and the jet in the higher transitions. Thus, the computed T$_{\rm rot}$ may represent an intermediate temperature between the jet and outflow and is therefore not used in section \ref{sec:outflow_jet} as either outflow or jet excitation temperature. 


\begin{deluxetable*}{l@{\extracolsep{2pt}}c@{\extracolsep{1pt}}c@{\hskip 1pt}c@{\extracolsep{1pt}}c@{\extracolsep{1pt}}c
}[h]
\tablecaption{H$_2$CO Line Properties\label{tab:H2CO_spatalogue}}
\tablewidth{0pt}
\tablehead{
\colhead{Transition} & \colhead{Frequency} & \colhead{$\log$(A$_{ul}$)} & \colhead{E$_u$}   & W & \colhead{Line}\\ 
\colhead{} & \colhead{(GHz)} &\colhead{(s$^{-1}$)}   &\colhead{(K)} & \colhead{(Jy\,bm$^{-1}$\,km\,s$^{-1}$)} & \colhead{}\\
}
\startdata
3( 0, 3)- 2( 0, 2) & 218.222192  & -3.55037 & 20.95640 & 0.084 & JPL\\
3( 2, 2)- 2( 2, 1) & 218.475632  & -3.80403 &  68.09370 & 0.028 & JPL\\
3( 2, 1)- 2( 2, 0) & 218.760066   & -3.80235 & 68.11081 & 0.024 & JPL\\
\enddata
\end{deluxetable*}

\subsection{Envelope Emission}\label{sec:envelope}

\begin{figure*}
\fig{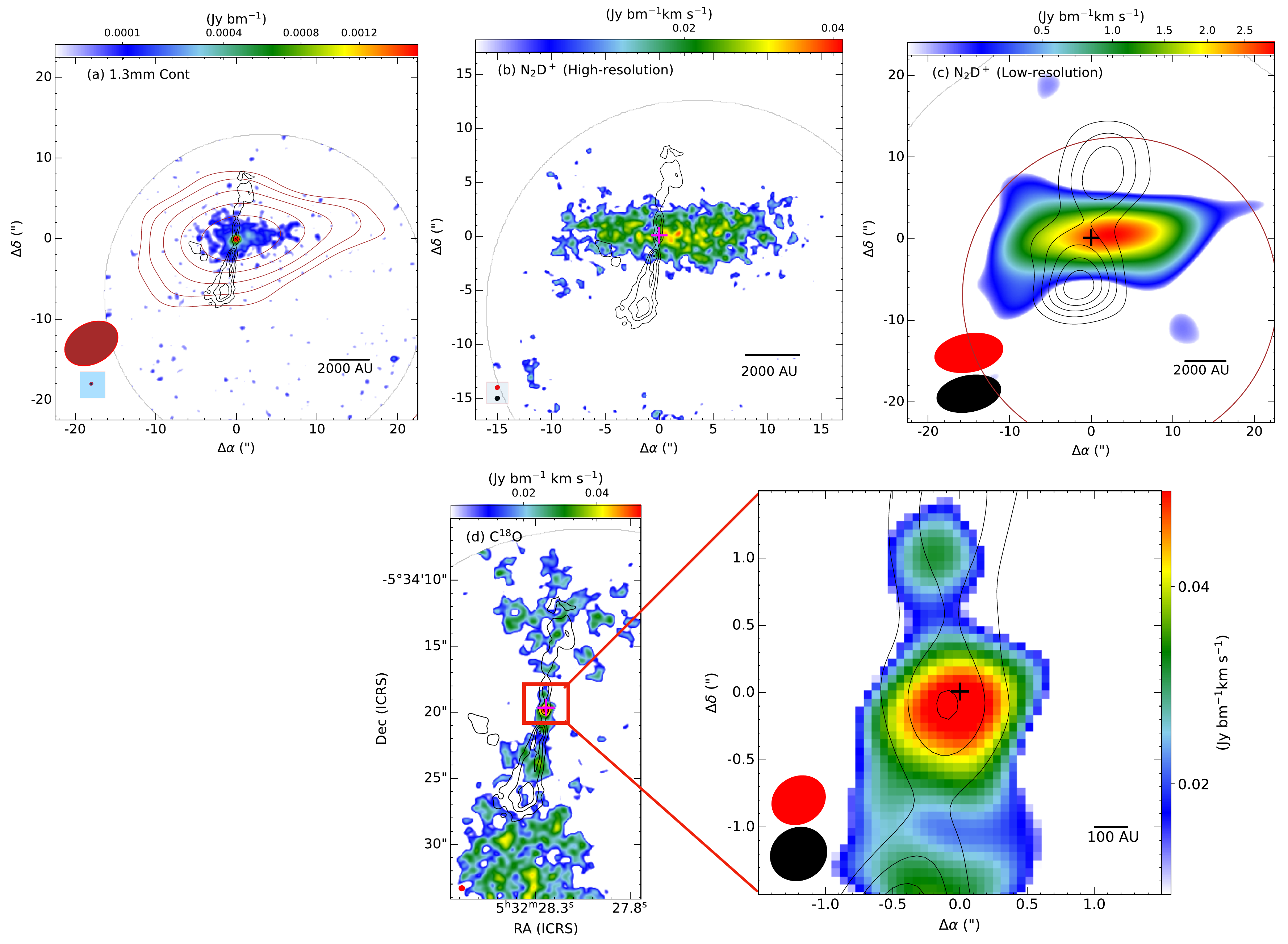}{0.98\textwidth}{}
\caption{(a) High-resolution ALMA 1.3\,mm continuum map with sensitivity of 0.035 mJy\,beam$^{-1}$ (beam-size in red in the lower left). Low-resolution 7-m ACA contours are over plotted in brown at 3$\times$ 2$^n$\ $\sigma$, where n = 1, 2, 3, 4, 5, 6 and sensitivity $\sigma$ = 0.4 mJy\,beam$^{-1}$ (beam-size in brown in the lower left). (b) High-resolution N$_2$D$^{+}$ map integrated over velocity 5.6 $-$ 7.0 km\,s$^{-1}$ with sensitivity of 10 mJy\,beam$^{-1}$km\,s$^{-1}$. 
(c) Low-resolution N$_2$D$^+$ map integrated over velocity 5.6--7.0 km\,s$^{-1}$ with sensitivity of 0.17 Jy\,beam$^{-1}$\,km\,s$^{-1}$. 
The CO contours of 7-m ACA observations are over plotted at 3$\times$(1, 2, 3, 4, 5, 6)$\sigma$, where $\sigma$ = 0.2 Jy\,beam$^{-1}$ km\,s$^{-1}$. The cross mark indicates the 1.3\,mm  continuum peak. The large circle indicates the combined TM1+TM2+ACA primary beam. 
(d)  High-resolution C$^{18}$O map integrated over velocity 7.0--8.4 km\,s$^{-1}$ with sensitivity  of 10 mJy\,beam$^{-1}$km\,s$^{-1}$. Central part is also shown in zoomed view. The CO contours have the same meaning of Figure \ref{fig:CO_SiO_PV}a. Beam sizes of CO (black) and C$^{18}$O emission map (red) are also shown in the lower left in each panel. The linear scale bars are shown in all the panels.}
\label{fig:cont_N2D_C18O}
\end{figure*}

Figure \ref{fig:cont_N2D_C18O}(a) shows the 1.3\,mm dust continuum emission. At  high-resolution, the continuum emission is resolved out, whereas at low-resolution (ACA) the emission is quite extended, revealing the source envelope. The observed line emission associated with the outflow is mostly confined within the envelope.
Gaussian fitting to the low-resolution ACA image provides a flux density (F$_\nu$) of 69.0 mJy, with a peak emission of 34.5 mJy\,beam$^{-1}$. The Gaussian fitted deconvolved parameters are tabulated in Table \ref{tab:Appendixcontinuum}. The core has an effective radius of R$_{\rm eff}$ = $\sqrt{\rm Maj*Min}$ $\sim$ 2500 AU. 

We derive the dust mass corresponding to this F$_\nu$ under the assumption of optically thin emission,  using the equation: 
\begin{equation}
M_{Env} \sim \frac{D^2 F_\nu}{B_\nu (T_{dust}) \kappa_\nu}
\end{equation}
where D is the distance to the Orion molecular cloud  $\sim$ 400 pc \citep{2018AJ....156...84K} and B$_\nu$ stands for  the Planck blackbody function at a dust temperature of T$_{dust}$.  Fitting the Planck blackbody function to the multiwavelength fluxes provided in Appendix \ref{sec:appendix_SED_2017}, we estimate T$_{dust}$ of 15 $\pm$ 5\,K. $\kappa_\nu$ represents the mass opacity of the protostellar core at 1.3\,mm, which can be expressed as $\kappa_\nu$ = 0.00899($\nu$/231 GHz)$^\beta$  cm$^2$ g$^{-1}$ \citep{2018ApJ...863...94L} in the early phase for coagulated dust particles with no ice mantles  \citep[see also, OH5: column 5 of][]{1994A&A...291..943O}, given a gas-to-dust mass ratio of 100 and spectral index $\beta$ $\sim$ 1.7 for the envelope. The final mass is estimated to be M$_{Env}$ $\sim$ 0.38 $\pm$ 0.14 M$_\sun$.

\begin{deluxetable}{c@{\extracolsep{1pt}}c@{\extracolsep{1pt}}
}[h]
\tablecaption{Deconvolved parameters of Gaussian fit at ACA continuum\label{tab:Appendixcontinuum}}
\tablewidth{0pt}
\tablehead{
\colhead{Deconvolved Parameters} & \colhead{value (Unit)}
}
\startdata
Flux & 69.0 $\pm$ 3.5 (mJy)\\
Peak & 34.5 $\pm$ 1.3 (mJy/bm)\\
Maj & 8.77 $\pm$ 0.77 ($\arcsec$)\\
Min & 4.45 $\pm$ 0.34 ($\arcsec$)\\
PA & 94.45 $\pm$ 4.75 ($\deg$)\\
Mass & 0.38 $\pm$ 0.14 ($M_\sun$)\\
R$_{\rm eff}$ & 2500 (AU)\\
\enddata
\end{deluxetable}

Deuterated species mainly trace the prestellar stage or very early phase of protostar evolution. Figure \ref{fig:cont_N2D_C18O}(b) and (c) delineate N$_2$D$^+$ emission at high and low-resolution, respectively.  Only two channels, at velocity 5.6 -- 7.0 km\,s$^{-1}$,  display N$_2$D$^+$ emission. The channel at 7.0 km\,s$^{-1}$ has the strongest emission and is considered the V$_{sys}$ for G208Walma.
G208Walma exhibits very extended N$_2$D$^{+}$ emission, 
which is quite similar in morphology to the 1.3\,mm continuum emission. Interestingly, the N$_2$D$^{+}$ peak is $\sim$ 500$\pm$300 AU apart from the dust continuum peak, estimated from Gaussian fitting to both images. Here, we assume the error bar to be of order twice the high-resolution beam size. This separation is evident in both the high-resolution as well as the low-resolution map. %
Assuming optically thin emission and an excitation temperature of 10\,K, we derive an average N$_2$D$^{+}$ column density N$_{\rm {N_2D^{+}}}$ $\sim$ 1.2 $\times$ 10$^{12}$ cm$^{-2}$.

C$^{18}$O emission, displayed in Figure \ref{fig:cont_N2D_C18O}(d), usually originates from the compact central part of the source. 
 Only two channels, at velocity 7.0 and 8.4 km\,s$^{-1}$, show C$^{18}$O emission. For this source, the C$^{18}$O emission is likely tracing the object as well as the ambient material. Given such complicated morphology, the true object emission is difficult to disentangle from our low-velocity resolution observations. Considering the uncertainties in estimation, we suggest that the C$^{18}$O emission in the central part (zoomed view of \ref{fig:cont_N2D_C18O}c) is possibly originating from a small central compact core where N$_2$D$^+$ is weak. This implies that the gas surrounding the central protostar is heated above the CO evaporation temperature ($\sim$ 20\,K).
Another interesting fact to note is that, comparing Figure \ref{fig:SiO_CO_H2CO_jet}a and \ref{fig:cont_N2D_C18O}c, the Southern lobe of the outflow tip is possibly interacting with ambient material. Thus, the SiO emission there 
may be due to collisional excitation.

\section{Discussion}\label{sec:discussion}

\begin{figure*}
\fig{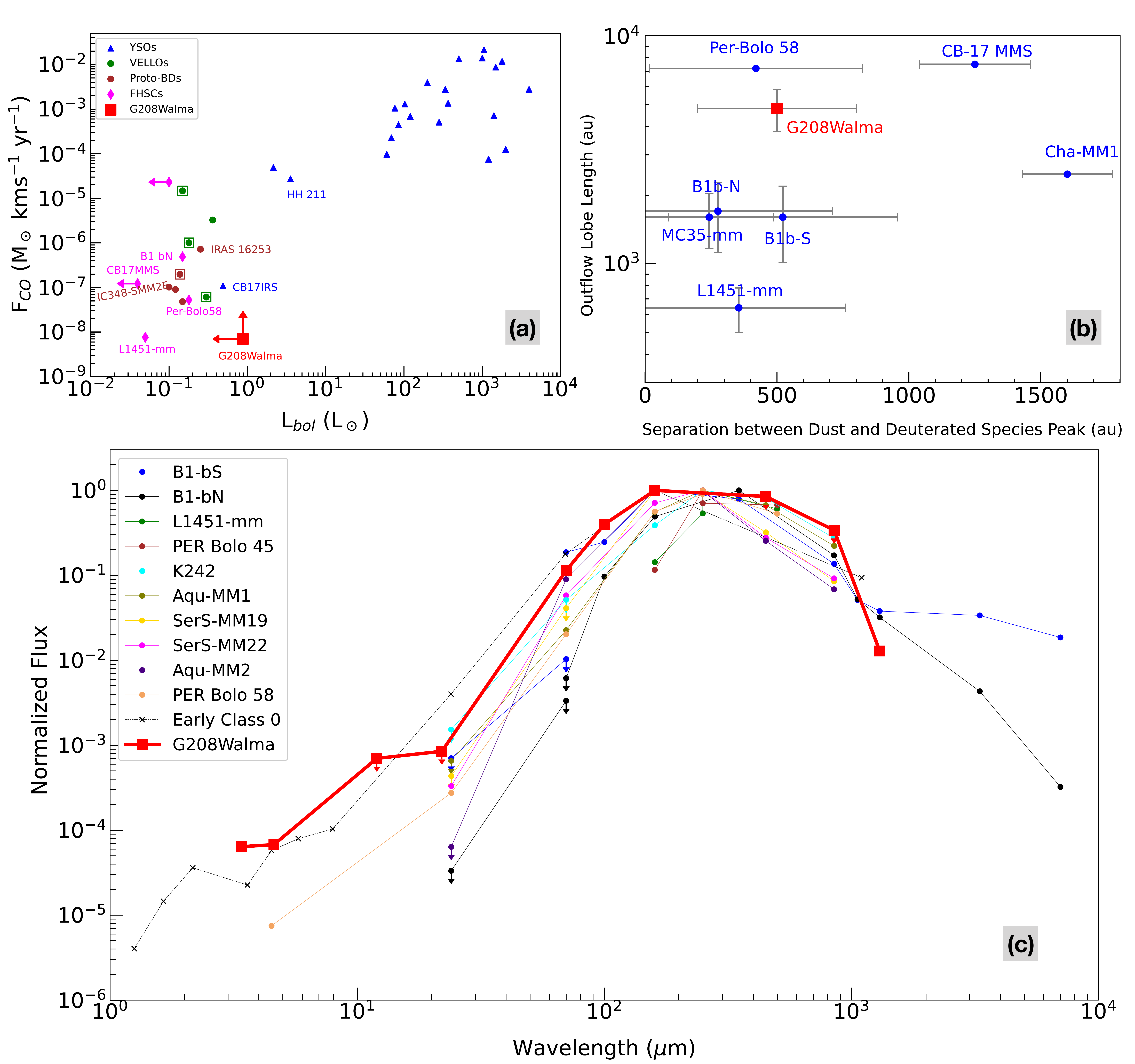}{0.98\textwidth}{}
\caption{(a) F$_{\rm CO}$ is plotted as a function of L$_{\rm bol}$. The figure is reproduced following Figure\,7 of \citet[][]{2014MNRAS.444..833P} and Figure\,13 of \citet[][]{2016ApJ...826...68H}. The red rectangle represents the G208Walma. Blue, green, brown, magenta points represent YSOs, VeLLOs, proto-BDs, candidate FHSCs. proto-BDs are also in the VeLLOs category. Open squares indicate single dish observations. The arrows attached to F$_{\rm CO}$ and L$_{\rm bol}$ for G208Walma indicate lower and upper limits, respectively. 
(b) CO outflow lobe length versus separation between dust continuum and deuterated species peaks for different candidate FHSCs and early protostars (marked with names). This figure is reproduced from Figure 14 (bottom panels) of \citet[][]{2020MNRAS.499.4394M}, and G208Walma is added by the red rectangle.
(c) Comparison of the normalized SED of G208Walma with literature candidate FHSCs and early protostars, following Figure 14 (top panel) of \citet[][]{2020MNRAS.499.4394M}. The corresponding multiwavelength fluxes (Table \ref{tab:Appendix_Multiwavelength_Flux}) and references (Table \ref{tab:Appendix_Multiwavelength_Flux_AllObjects}) are presented in Appendix \ref{sec:appendix_SED_2017}.  Different colors represent different objects and average early Class\,0 SED (T$_{bol}$ $<$ 50\,K). 
G208Walma is shown as a red line with rectanglar points. 
}
\label{fig:normalized_sed_comparison}
\end{figure*}

A small observed outflow velocity, V$_{\rm obs,out}$  $\lesssim$ 10 km\,s$^{-1}$  is one of the main characteristics of candidate FHSCs \citep[][]{2014ApJ...789...50H,2020MNRAS.499.4394M}. G208Walma exhibits  V$_{\rm obs,out}$ $\sim$ 4.2 km\,s$^{-1}$, analogous to those of FHSCs. However, the deprojected V$_{\rm j}$ $\sim$ 24.0 km\,s$^{-1}$ is higher than that of the FHSC limit.  The outflow is spatially extended up to 4800 AU when deprojected and has a t$_{\rm dyn}$ $\sim$ 930$^{+200}_{-100}$ years, which suggests that  the object could be at a very early phase. The outflow lobes appear to be collimated in the high resolution maps (Figure \ref{fig:SiO_CO_H2CO_jet}a-d), similar to Class\,0 protostars.  %

A correlation between F$_{\rm CO}$ and L$_{\rm bol}$  has been investigated in the literature to probe the evolutionary stages of protostars \citep[e.g.,][]{1996A&A...311..858B}. 
Using multiwavelength observed flux, we estimate the bolometric luminosity (L$_{\rm bol}$) and bolometric temperature (T$_{\rm bol}$) for G208Walma (see Appendix \ref{sec:appendix_SED_2017} for the details). In Figure \ref{fig:normalized_sed_comparison}a, we compare G208Walma with normal young stellar objects (YSOs), very low-luminosity objects (VeLLOs), proto brown dwarfs (proto-BDs), and candidate FHSCs  from  \citet[][see their Table\,4 and Figure\,7, and references therein]{2014MNRAS.444..833P} and \citet[][see their Figure\,13]{2016ApJ...826...68H}. Here we note that proto-BDs are also in the VeLLOs category.
  When the protostar evolves from the initial collapse phase, the luminosity of the  central core increases. The accretion/ejection activity is also expected to increase up to a certain phase, possibly up to the late Class\,0 or early Class\,I, consequently the F$_{\rm CO}$ should increase with protostellar evolution. Therefore, the YSOs in Figure \ref{fig:normalized_sed_comparison}a exhibit higher L$_{\rm bol}$ and higher F$_{\rm CO}$ compared with the youngest protostars and candidate FHSCs, VeLLOs, proto-BDs.

G208Walma appears much younger than YSOs like HH\,211, and shares a lower limit of F$_{\rm CO}$ and an upper limit for L$_{\rm bol}$ with the VeLLOs, Proto-BDs and FHSCs in Figure \ref{fig:normalized_sed_comparison}a. It is likely more evolved than the candidate FHSCs L1451-mm. G208Walma is embedded within a large reservoir (effective radius $\sim$ 16,500 au) with mass $\sim$ 4.78 M$_\sun$, estimated from the 850 $\mu$m emission obtained by the James Clerk Maxwell Telescope (JCMT) \citep[island \#25 in][]{2016MNRAS.461.4022M}. From the 1.3\,mm continuum emission, it possesses a thick envelope of mass $>$ 0.38 M$_\sun$. If we assume 1/3 of the envelope would be accreted to the central protostar then  even with the mass observed with ALMA, a star of mass $>$ 0.12 M$_\sun$ will form. Thus, G208Walma may not form a proto-BDs but rather a very low-mass star, and it may have already passed the FHSC phase or be transiting from an FHSC to a protostar. 

 Detection of SiO in a jet signifies very high accretion/ejection activity within a protostar.  The SiO jet is usually launched from the earliest Class\,0 protostars having very high jet mass-loss (\.{M}$_j$  $\gtrsim$ 10$^{-6}$ M$_\sun$\,yr$^{-1}$) and very high V$_j$ ($>$ 100 km\,s$^{-1}$) e.g., B\,335 \citep[][]{2019A&A...631A..64B},
HH\,211 \citep[][]{2018ApJ...863...94L}. 
Dense SiO emission is detected within the outflow cavity of G208Walma, although it exhibits a smaller V$_j$ and \.{M}$_j$ ($\sim$ 1.96 $\times$ 10$^{-7}$ M$_\sun$\,yr$^{-1}$; Table \ref{tab:jet_outflow}), indicating that a protostar may have already formed inside the core which has passed through the second collapse phase or is transitioning from an FHSC to a protostar. To form SiO in the jet, the SiO needs to be synthesized from larger dust grains either (i) at the shock region in the jet through grain sputtering or (ii) at the dust sublimation zone near the protostellar core. In the first case, the shock velocity (V$_s$) is predicted to be 10 $<$ V$_s$ $<$ 40 km\,s$^{-1}$ \citep[][]{1997A&A...321..293S}. For G208Walma, the observed V$_s$ is $\sim$ 4 km\,s$^{-1}$ in the Northern lobe, whereas the southern part is blended with ambient material and difficult to measure (Figure \ref{fig:CO_SiO_PV}b). The V$_s$ could be affected largely due to our low sensitivity and low velocity resolution.
For the latter case, the Si could have been released from grain surfaces in  the dust sublimation zone near the object and synthesized into SiO. The L$_{bol}$ of 0.80 L$_\sun$ suggests that the dust sublimation radius could be $<$ 0.1 AU  \citep[][]{2007prpl.conf..539M}. Comparison of the measured X[SiO/CO] and \.{M}$_j$ from Table \ref{tab:jet_outflow} with the  astrochemical model of \citet[][their Figure 12]{2020A&A...636A..60T} suggests that the jet may have been launched from the outermost region of the dust-free zone, where the dust-to-gas ratio (Q) is relatively higher (0.1 to 0.01). 

 \citet[][]{2012ApJ...745...18S} detected extended SiO (2–1) emission from another very young source, Per-Bolo\,45, at a distance of 20$\arcsec$ south of the object and with velocities within 1 km\,s$^{-1}$ of the ambient velocity. They interpreted this detection as tentative evidence for a jet in  Per-Bolo\,45. However, \citet[][]{2020MNRAS.499.4394M} extensively studied this object and suggested that this SiO (2-1) emission is possibly not associated with Per-Bolo\,45, rather it is a collision zone between Per-Bolo\,45 and the outflow tip of another source SVS\,13c. They also concluded that Per-Bolo\,45 is pre-stellar in nature where a compact object  is not formed yet, such as a case of an FHSC.

As starless cores evolve toward collapse, the molecular deuterium fraction in the gas phase (e.g., N$_2$D$^+$/N$_2$H$^+$) increases and reaches maximum at the onset of star formation \citep[][]{2005ApJ...619..379C,2009A&A...493...89E,2021ApJS..256...25T}. Once the protostar is formed inside the core, the deuterium fraction declines 
\citep[][]{2015A&A...579A..80G,2021arXiv211113325S} as the star heats up the surrounding medium and dissociates deuterated species. As a consequence, the centre of the observed N$_2$D$^+$ peak could be shifted away from the protostar, 
whereas the dust continuum peak remains unaltered and represents the envelope emission around the central protostar. Therefore, the shift in N$_2$D$^+$ emission and dust continuum peak could provide a possible indication of the evolution of the dense core. 
In the case of G208Walma, the continuum peak is observed at the middle of the dense core (in Figure \ref{fig:cont_N2D_C18O}c) and the approximate size of the dense core on both sides of the continuum peak appears the same. The N$_2$D$^+$ peak, however, is shifted by $\sim$ 500$\pm$300 AU. We compare this separation and outflow lobe length with other young known FHSC candidates and young protostars in Figure \ref{fig:normalized_sed_comparison}b. G208Walma appears to be more evolved than the most promising candidate FHSC L1451-mm \citep[][]{2020MNRAS.499.4394M}, and comparable with the young protostar/candidate FHSC B1b-S \citep[][]{2014ApJ...789...50H}. It might also be younger than Cha-MM1, CB-17\,MMS \citep[][]{2012ApJ...751...89C}. We caution, however, that high uncertainties in the measured separation prevent drawing a specific evolutionary trend. %

The estimated  T$_{\rm bol}$ and L$_{\rm bol}$ for G208Walma (Appendix \ref{sec:appendix_SED_2017}) suggest that a protostar may have already formed inside the core, although it is not much evolved towards Class\,0 phase. In Figure \ref{fig:normalized_sed_comparison}c, we compare the SED  of G208Walma with other known candidate FHSCs and average early Class\,0 spectra (T$_{\rm bol}$ $<$ 50\,K) from \citet[][]{2009ApJ...692..973E}. From 24 $\mu$m and above, G208Walma spectra fairly resembles to the candidate FHSCs with the SED peak is around 150-250 $\mu$m. %

\section{Conclusions}
In this work we analyze ALMA observations at 1.3\,mm continuum and molecular line emission for the object G208Walma. The observations suggest a low-velocity outflow from G208Walma is present (observed velocity $\sim$ 4.2 km\,s$^{-1}$ and corresponding deprojected V$_{\rm j}$ $\sim$ 24.0 km\,s$^{-1}$), with compact SiO emission along the outflow suggesting the presence of a dense jet. We estimate a smaller jet mass-loss rate (\.{M}$_j$ $\sim$ 1.96 $\times$ 10$^{-7}$ M$_\sun$\,yr$^{-1}$) and smaller outflow force (F$_{\rm CO}$ $\sim$ 0.8 $\times$ 10$^{-8}$  M$_\sun$km\,s$^{-1}$/yr) than that of other observed protostars in the literature. Dynamical time (t$_{dyn}$ $\sim$ 930$^{+200}_{-100}$ years), extended 1.3\,mm emission (R$_{\rm eff}$ $\sim$ 2500 AU), extended and offset N$_2$D$^+$ emission,  and the SED suggest that the source is at the very early stage of protostellar evolution, possibly within a few hundred years of the second collapse. Therefore, G208Walma could be the earliest protostellar object  with a molecular jet observed to date.\\



\begin{acknowledgments}
This paper makes use of the following ALMA data:  ADS/JAO.ALMA$\#$2018.1.00302.S. ALMA is a partnership of ESO (representing its member states), NSF (USA) and NINS (Japan), together with NRC (Canada), NSC and ASIAA (Taiwan), and KASI (Republic of Korea), in cooperation with the Republic of Chile. The Joint ALMA Observatory is operated by ESO, AUI/NRAO and NAOJ.  S.D. and C.-F.L. acknowledge grants from the Ministry of Science and Technology of Taiwan (MoST 107-2119-M- 001- 040-MY3) and the Academia Sinica (Investigator Award AS-IA-108-M01). Tie Liu acknowledges the supports by National Natural Science Foundation of China (NSFC) through grants No.12073061 and No.12122307, the international partnership program of Chinese Academy of Sciences through grant No.114231KYSB20200009, Shanghai Pujiang Program 20PJ1415500 and the science research grants from the China Manned Space Project with no. CMS-CSST-2021-B06. This research was carried out in part at the Jet Propulsion Laboratory, which is operated by the California Institute of Technology under a contract with the National Aeronautics and Space Administration (80NM0018D0004). C.W.L. is supported by the Basic Science Research Program through the National Research Foundation of Korea (NRF) funded by the Ministry of Education, Science and Technology (NRF-2019R1A2C1010851).  D.J. is supported by the National Research Council of Canada and by a Natural Sciences and Engineering Research Council of Canada (NSERC) Discovery Grant. S.L. Qin is supported by the National Natural Science Foundation of China under grant No. 12033005. P.S. was partially supported by a Grant-in-Aid for Scientific Research (KAKENHI Number 18H01259) of the Japan Society for the Promotion of Science (JSPS). LB gratefully acknowledges support by the ANID BASAL projects ACE210002 and FB21000. VMP acknowledges support by the grant PID2020-115892GB-I00 funded by MCIN/AEI/10.13039/501100011033.
\end{acknowledgments}

\facility{ALMA}\\

\software{Astropy \citep[][]{2013A&A...558A..33A}, APLpy \citep[][]{2012ascl.soft08017R}, Matplotlib \citep[][]{Hunter:2007}, CASA \citep[][]{2007ASPC..376..127M}}



\newpage
\appendix

\setcounter{figure}{0} 
\renewcommand{\thefigure}{A\arabic{figure}} 
\begin{figure*}
\fig{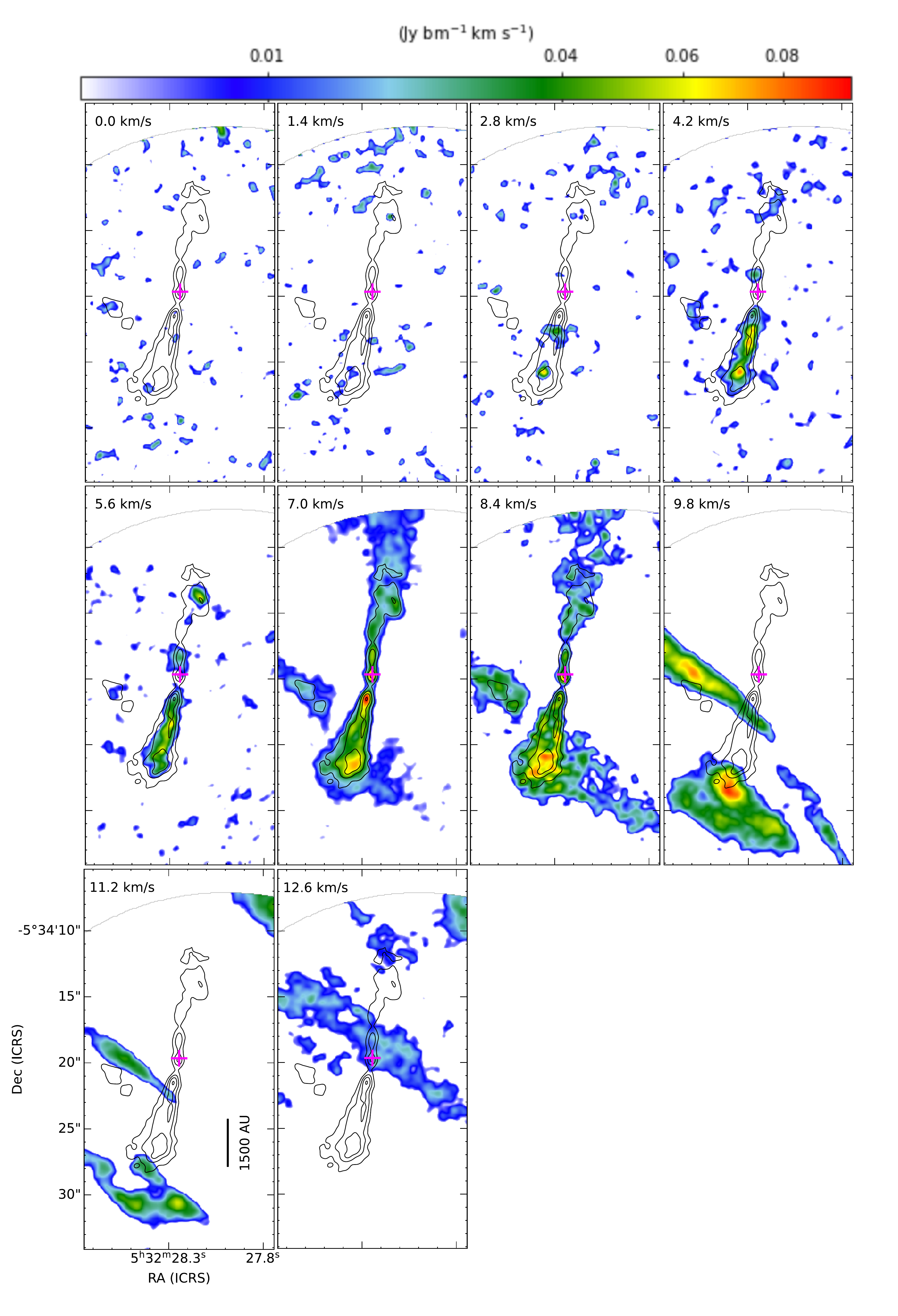}{0.85\textwidth}{}
\caption{
Channel maps of $^{12}$CO(2$-$1) emission in color scale. The black contours represent the integrated CO emission, and have the same meaning as Figure \ref{fig:CO_SiO_PV}a. The corresponding channel velocities are mentioned in each panel. The magenta cross is the 1.3\,mm continuum peak.  A linear scale bar is shown in the lower left panel (at channel V$_{\rm obs}$ = 11.2 km s$^{-1}$)}.
\label{fig:Appendix_CO_channelMap}
\end{figure*}


\section{CO  emission and Jet/Outflow parameters}\label{sec:Appendix_CO_properties}
\subsection{Channel Maps and Spectra}\label{sec:Appendix_CO_channelMap}
Figure \ref{fig:Appendix_CO_channelMap} displays the CO channel maps at different velocities. The outflow emission channels in the Southern lobe range from V$_{\rm obs}$ = 2.8 to 8.4 km\,s$^{-1}$. The channel at V$_{\rm obs}$ $\sim$ 8.4 km\,s$^{-1}$ is possibly showing both outflow and ambient material. The Northern channels range from V$_{\rm obs}$ = 4.2 to 8.4 km\,s$^{-1}$, where the channel at V$_{\rm obs}$ = 2.8 km\,s$^{-1}$ also shows some faint emission. The systemic velocity of G208Walma is V$_{sys}$ = 7.0 km\,s$^{-1}$. For both the lobes, the maximum flow velocities are considered to be V$_{\rm off}$ = $|V_{\rm obs}$ -- $V_{\rm sys}|$ = 4.2 km\,s$^{-1}$. The outflow emission in the southern and northern velocity channels are not equally distributed on both side of V$_{sys}$, therefore, it is difficult to define blueshifted and redshifted emission precisely for G208Walma.
Such outflow velocity structure around V$_{\rm sys}$ suggests a small inclination angle of the system.  
The channels from V$_{\rm obs}$ = 7.0 to 11.2 km\,s$^{-1}$ indicate that the tip of the southern jet/outflow lobe is possibly colliding with ambient material. Thus, the SiO emission at the outermost part of the Southern lobe could be a combination of jet emission and a collision zone.

\begin{figure}
\fig{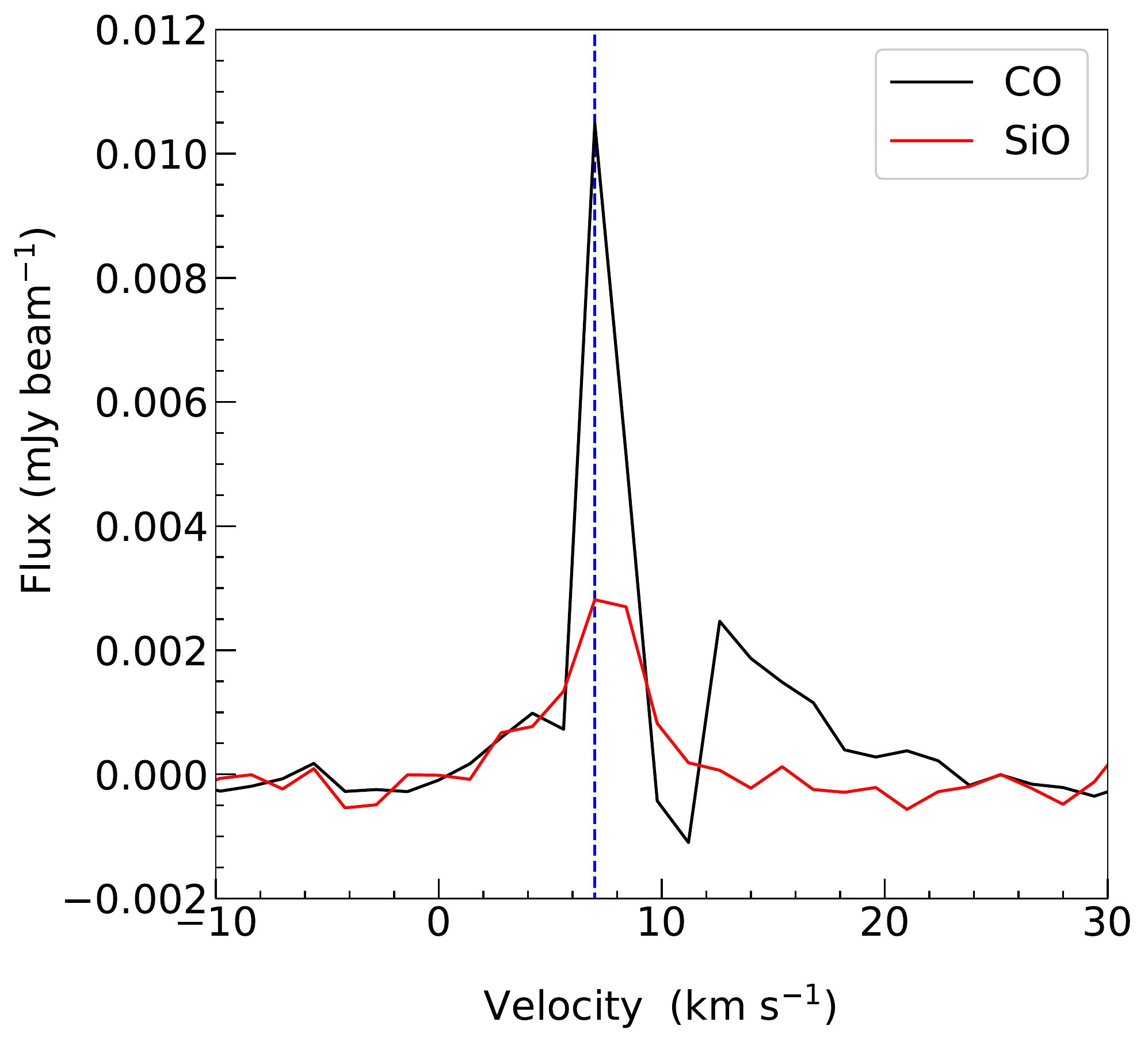}{0.45\textwidth}{}
\caption{
 Spectra CO (in black) and SiO (in red) emission extracted from a box region along an outflow/jet axis, which covers total CO and SiO emission, respectively. The systemic velocity at V$_{sys}$ = 7.0 km\,s$^{-1}$ is marked with vertical blue dashed line. 
}
\label{fig:Appendix_CO_SiO_spectra}
\end{figure}

 Spectra in CO and SiO  emission from a rectangular region along the outflow/jet axis are displayed in Figure \ref{fig:Appendix_CO_SiO_spectra}. The SiO emission is confined between 2.8 to 10 km\,s$^{-1}$. The CO emission has two peaks: one at  V$_{sys}$ $\sim$ 7.0 km\,s$^{-1}$ and another at $\sim$ 12.0-14.0 km\,s$^{-1}$. From channel maps in Figure \ref{fig:Appendix_CO_channelMap}, it is obvious that there is no counterpart in the CO outflow at velocity 9.8 km\,s$^{-1}$ and above. The SiO spectra also confirmed that the second peak at $\sim$ 12.0-14.0 km\,s$^{-1}$ is not associated with G208Walma. In the spectra of CO and SiO, there is no clear differentiation between the high-velocity and low-velocity component. Therefore, it is difficult to distinguish between the low-velocity CO outflow (wind component) and the high-velocity jet component based on the velocity distribution only.

\subsection{Jet mass-loss rate}\label{sec:appendix_results_jetmassloss_rate}
The jet mass-loss rate \.{M$_j$} was derived from the CO emission flux for a specific excitation temperature of T$_{ex}$ within the jet. Under the assumption of optical thin CO emission, the beam-average CO column densities (N$_{CO}$) were measured, which is then converted into beam-averaged H$_2$ column density N$_{H_2}$ for CO abundance ratio, X$_{CO}$  = N$_{CO}$/N$_{H_2}$. We assume that the molecular jet is flowing through a uniform cylinder at a constant density and speed of gas along the transverse beam direction. Since the jet is not resolved at the present spatial resolution, the beam size ($b_m$) is taken to be the jet width.  Thus, \.{M}$_j$ can be expressed as:
\begin{equation}\label{equ:mass_equation}
\indent
\dot{M_j} = \mu_{H_2} m_H \frac{N_{CO}}{X_{CO}}\, V_{j}\, b_m,\\\\
\end{equation}
where $\mu_{H2}$ = 2.8 is the mean molecular weight and $m_H$ is the mass of a  hydrogen atom. V$_{j}$ is the mean deprojected jet velocity.

\subsection{Outflow force}\label{sec:Appendix_result_forcemomentum}

We derive the outflow force (F$_{CO}$) from CO the emission. First, the outflow emission above 3$\sigma$ in the k-th channel is converted to outflow mass (M$_{k}$) for each velocity channel following the equation \citep[][]{2015A&A...576A.109Y}:
\begin{equation}\label{equ:outflowmass_equation}
\indent
M_{k} = \mu_{H_2} m_H A \frac{\sum_{l}{N_{CO,l}}}{X_{CO}}\\\\
\end{equation}
where the sum is over all outflow pixels $l$ on the k-th channel. N$_{CO}$ is the  beam-average CO column density. A is the surface area of each pixel and CO abundance ratio of X$_{CO}$ $\sim$ 10$^{-4}$ in the jet. In the next step, for the outflow extension of R$_{CO}$ and maximum outflow velocity (V$_{CO,max}$), F$_{CO}$ can be expressed as 
\citep[][]{2015A&A...576A.109Y}:
\begin{equation}\label{equ:force_equation}
F_{CO} = f_{ia}\frac{V_{CO,max}\sum_k{M_k V_k}}{R_{CO}}
%
\end{equation}
where the momentum ($P$) on k-th channel is M$_k$V$_k$ for a mass M$_k$ with the central velocity, V$_k$ (= $|$\,V$_{\rm obs}$ $-$ V$_{\rm sys}$\,$|$). The factor $f_{ia}$ deals with the inclination correction. 

%

\setcounter{table}{0} 
\renewcommand{\thetable}{B\arabic{table}} 
\section{Spectral Energy Distribution}\label{sec:appendix_SED_2017}
We searched published catalogues for a compact continuum counterpart to G208Walma within a search radius of 3$\arcsec$ from the 1.3 continuum peak \citep[see][for more details on catalog matching method]{2015MNRAS.454.3597D}. The  Wide-field Infrared Survey Explorer \citep[WISE;][]{2010AJ....140.1868W} and  Herschel Space Observatory (https://sci.esa.int/web/herschel) data were obtained from IRSA IPAC catalog  (https://irsa.ipac.caltech.edu/). It matches with WISE catalogue source J053228.14-053420.6. James Clerk Maxwell Telescope (JCMT) data were obtained from \citet[][]{2016MNRAS.461.4022M} (island \#25). 
The multiwavelength fluxes are listed in Table \ref{tab:Appendix_Multiwavelength_Flux}.
Based on the poor signal-to-noise ratio, the fluxes at WISE 12 $\mu$m, 24 $\mu$m and JCMT 450$\mu$m, 850 $\mu$m are considered as an upper limit measurement. 

T$_{bol}$ and L$_{bol}$ were derived using  trapezoid-rule of integration over all the observed fluxes. Following \cite{1993ApJ...413L..47M}, the flux weighted mean frequencies in the observed SED were used to estimate T$_{bol}$.  We obtained T$_{bol}$ $\sim$ 31 K and L$_{bol}$ $\sim$ 0.8 L$_\sun$. Based on the uncertainty in the observed fluxes, we assume these values are at the upper limit.

\begin{deluxetable}{c@{\extracolsep{1.0cm}}c@{\extracolsep{1.0cm}}c@{\extracolsep{1.0cm}}c@{\extracolsep{1pt}}
}[h]
\tablecaption{Multiwavelength Flux of G208Walma\label{tab:Appendix_Multiwavelength_Flux}}
\tablewidth{0pt}
\tablehead{
\colhead{Wavelength} & \colhead{Flux} & \colhead{Error} & \colhead{Reference}\\
\colhead{($\mu$m)} & \colhead{(mJy)} &\colhead{(mJy))}   &\colhead{Telescope} 
}
\startdata
3.4 &  0.3435 & 0.0261 & WISE \\
4.6 & 0.3626 & 0.0256 & WISE \\
12.0 & 3.7558 & 1.2804 & WISE \\
22.0 & 4.45537 & $--$ & WISE\\
70.0 & 609.079 & 13.356 & Herschel\\
100.0 & 2139.573 & 58.847 & Herschel\\
160.0 & 5359.623 & 17.7 & Herschel\\
450.0 & 4540.0 &  $--$  & JCMT\\
850.0 & 1820.0 &   $--$  & JCMT\\
1300.0 & 69.0 & 3.5 & ALMA; This study\\
\enddata
\end{deluxetable}

\begin{deluxetable}{c@{\extracolsep{10pt}}c@{\extracolsep{1pt}}
}[h]
\tablecaption{References for multiwavelength fluxes of Figure \ref{fig:normalized_sed_comparison}c
\label{tab:Appendix_Multiwavelength_Flux_AllObjects}}
\tablewidth{0pt}
\tablehead{
\colhead{Sources} & \colhead{Catalog reference}
}
\startdata
B1-bN & \citet[][]{2014ApJ...789...50H} \\
B1-bS & \citet[][]{2014ApJ...789...50H} \\
L1451-mm & \citet[][]{2020MNRAS.499.4394M} \\  
PER Bolo 45 & \citet[][]{2020MNRAS.499.4394M} \\
K242 & \citet[][]{2018MNRAS.474..800Y} \\
Aqu-MM1 & \citet[][]{2018MNRAS.474..800Y} \\
SerpS-MM19 & \citet[][]{2018MNRAS.474..800Y} \\
SerpS-MM22 & \citet[][]{2018MNRAS.474..800Y} \\
Aqu-MM2 & \citet[][]{2018MNRAS.474..800Y} \\
PER Bolo 58 & \citet[][]{2010ApJ...722L..33E,2020MNRAS.499.4394M} \\
Average Early Class 0 & \citet[][]{2009ApJ...692..973E} \\
\enddata
\end{deluxetable}


\bibliography{sample631}{}
\bibliographystyle{aasjournal}

\end{document}